\pdfoutput=1

\documentclass[12pt,a4paper]{article}

\usepackage{ifthen} 
\newboolean{pdflatex}
\setboolean{pdflatex}{true} 

\newboolean{articletitles}
\setboolean{articletitles}{true} 

\newboolean{uprightparticles}
\setboolean{uprightparticles}{false} 

\newboolean{inbibliography}
\setboolean{inbibliography}{false} 

\usepackage{microtype}
\usepackage{lineno}  
\usepackage{xspace} 
\usepackage{booktabs}

\usepackage{graphicx}  
\usepackage{color}
\usepackage{colortbl}
\graphicspath{{./figs/}} 

\usepackage{amsmath} 
\usepackage{amssymb}
\usepackage{amsfonts}
\usepackage{upgreek} 

\newcommand*\patchAmsMathEnvironmentForLineno[1]{%
\expandafter\let\csname old#1\expandafter\endcsname\csname #1\endcsname
\expandafter\let\csname oldend#1\expandafter\endcsname\csname
end#1\endcsname
 \renewenvironment{#1}%
   {\linenomath\csname old#1\endcsname}%
   {\csname oldend#1\endcsname\endlinenomath}%
}
\newcommand*\patchBothAmsMathEnvironmentsForLineno[1]{%
  \patchAmsMathEnvironmentForLineno{#1}%
  \patchAmsMathEnvironmentForLineno{#1*}%
}
\AtBeginDocument{%
\patchBothAmsMathEnvironmentsForLineno{equation}%
\patchBothAmsMathEnvironmentsForLineno{align}%
\patchBothAmsMathEnvironmentsForLineno{flalign}%
\patchBothAmsMathEnvironmentsForLineno{alignat}%
\patchBothAmsMathEnvironmentsForLineno{gather}%
\patchBothAmsMathEnvironmentsForLineno{multline}%
}

\usepackage{hyperref}    
\usepackage[all]{hypcap} 

\def\lhcb {\mbox{LHCb}\xspace}
\def\ux85 {\mbox{UX85}\xspace}

\def\belle  {\mbox{Belle}\xspace}

\ifthenelse{\boolean{uprightparticles}}%
{

 \def\Ppi         {\ensuremath{\uppi}\xspace}

 \def\Ppsi        {\ensuremath{\uppsi}\xspace}

 \def\PDelta      {\ensuremath{\Delta}\xspace}                 
 \def\PXi      {\ensuremath{\Xi}\xspace}                 
 \def\PLambda      {\ensuremath{\Lambda}\xspace}                 
 \def\PSigma      {\ensuremath{\Sigma}\xspace}                 
 \def\POmega      {\ensuremath{\Omega}\xspace}                 
 \def\PUpsilon      {\ensuremath{\Upsilon}\xspace}                 
 
 \def\PB      {\ensuremath{\mathrm{B}}\xspace}                 
                  
 \def\PD      {\ensuremath{\mathrm{D}}\xspace}

 \def\PJ      {\ensuremath{\mathrm{J}}\xspace}                 
 \def\PK      {\ensuremath{\mathrm{K}}\xspace}

 \def\Pb      {\ensuremath{\mathrm{b}}\xspace}                 
 \def\Pc      {\ensuremath{\mathrm{c}}\xspace}

 \def\Pi      {\ensuremath{\mathrm{i}}\xspace}

}
{

 \def\Ppi         {\ensuremath{\pi}\xspace}

 \def\Ppsi        {\ensuremath{\psi}\xspace}                 
                  
 \mathchardef\PDelta="7101
 \mathchardef\PXi="7104
 \mathchardef\PLambda="7103
 \mathchardef\PSigma="7106
 \mathchardef\POmega="710A
 \mathchardef\PUpsilon="7107
                  
 \def\PB      {\ensuremath{B}\xspace}                 
                  
 \def\PD      {\ensuremath{D}\xspace}

 \def\PJ      {\ensuremath{J}\xspace}                 
 \def\PK      {\ensuremath{K}\xspace}

 \def\Pb      {\ensuremath{b}\xspace}                 
 \def\Pc      {\ensuremath{c}\xspace}

 \def\Pi      {\ensuremath{i}\xspace}

}

\def\cquark    {\ensuremath{\Pc}\xspace}

\def\bquark    {\ensuremath{\Pb}\xspace}
\def\bquarkbar {\ensuremath{\overline \bquark}\xspace}
\def\bbbar     {\ensuremath{\bquark\bquarkbar}\xspace}

\def\pion  {\ensuremath{\Ppi}\xspace}

\def\kaon  {\ensuremath{\PK}\xspace}
  \def\Kbar  {\kern 0.2em\overline{\kern -0.2em \PK}{}\xspace}

\def\Kz    {\ensuremath{\kaon^0}\xspace}
\def\Kzb   {\ensuremath{\Kbar^0}\xspace}
\def\KzKzb {\ensuremath{\Kz \kern -0.16em \Kzb}\xspace}
\def\Kp    {\ensuremath{\kaon^+}\xspace}
\def\Km    {\ensuremath{\kaon^-}\xspace}

\def\KpKm  {\ensuremath{\Kp \kern -0.16em \Km}\xspace}

  \def\Dbar    {\kern 0.2em\overline{\kern -0.2em \PD}{}\xspace}
\def\D       {\ensuremath{\PD}\xspace}

\def\Dz      {\ensuremath{\D^0}\xspace}
\def\Dzb     {\ensuremath{\Dbar^0}\xspace}
\def\DzDzb   {\ensuremath{\Dz {\kern -0.16em \Dzb}}\xspace}
\def\Dp      {\ensuremath{\D^+}\xspace}
\def\Dm      {\ensuremath{\D^-}\xspace}

\def\DpDm    {\ensuremath{\Dp {\kern -0.16em \Dm}}\xspace}

  \def\Bbar    {\kern 0.18em\overline{\kern -0.18em \PB}{}\xspace}

\def\jpsi     {\ensuremath{{\PJ\mskip -3mu/\mskip -2mu\Ppsi\mskip 2mu}}\xspace}

  \def\Y#1S{\ensuremath{\PUpsilon{(#1S)}}\xspace}

\def\Lbar {\ensuremath{\kern 0.1em\overline{\kern -0.1em\PLambda}}\xspace}

\def\to                 {\ensuremath{\rightarrow}\xspace}

\def\AT#1     {\ensuremath{A_{\mathrm{T}}^{#1}}\xspace}           

\def\C#1      {\ensuremath{\mathcal{C}_{#1}}\xspace}                       
\def\Cp#1     {\ensuremath{\mathcal{C}_{#1}^{'}}\xspace}                    
\def\Ceff#1   {\ensuremath{\mathcal{C}_{#1}^{\mathrm{(eff)}}}\xspace}        
\def\Cpeff#1  {\ensuremath{\mathcal{C}_{#1}^{'\mathrm{(eff)}}}\xspace}       
\def\Ope#1    {\ensuremath{\mathcal{O}_{#1}}\xspace}                       
\def\Opep#1   {\ensuremath{\mathcal{O}_{#1}^{'}}\xspace}                    

\newcommand{\tev}{\ensuremath{\mathrm{\,Te\kern -0.1em V}}\xspace}
\newcommand{\gev}{\ensuremath{\mathrm{\,Ge\kern -0.1em V}}\xspace}
\newcommand{\mev}{\ensuremath{\mathrm{\,Me\kern -0.1em V}}\xspace}
\newcommand{\kev}{\ensuremath{\mathrm{\,ke\kern -0.1em V}}\xspace}
\newcommand{\ev}{\ensuremath{\mathrm{\,e\kern -0.1em V}}\xspace}
\newcommand{\gevc}{\ensuremath{{\mathrm{\,Ge\kern -0.1em V\!/}c}}\xspace}
\newcommand{\mevc}{\ensuremath{{\mathrm{\,Me\kern -0.1em V\!/}c}}\xspace}
\newcommand{\gevcc}{\ensuremath{{\mathrm{\,Ge\kern -0.1em V\!/}c^2}}\xspace}
\newcommand{\gevgevcccc}{\ensuremath{{\mathrm{\,Ge\kern -0.1em V^2\!/}c^4}}\xspace}
\newcommand{\mevcc}{\ensuremath{{\mathrm{\,Me\kern -0.1em V\!/}c^2}}\xspace}

\def\mum  {\ensuremath{\,\upmu\rm m}\xspace}

\def\mub{\ensuremath{\rm \,\upmu b}\xspace}

\def\invpb {\ensuremath{\mbox{\,pb}^{-1}}\xspace}

\def\invfb   {\ensuremath{\mbox{\,fb}^{-1}}\xspace}

\def\gsim{{~\raise.15em\hbox{$>$}\kern-.85em
          \lower.35em\hbox{$\sim$}~}\xspace}
\def\lsim{{~\raise.15em\hbox{$<$}\kern-.85em
          \lower.35em\hbox{$\sim$}~}\xspace}

\def\dllkpi     {\ensuremath{\mathrm{DLL}_{\kaon\pion}}\xspace}

\def\evtgen     {\mbox{\textsc{EvtGen}}\xspace}
\def\pythia     {\mbox{\textsc{Pythia}}\xspace}

\def\geant      {\mbox{\textsc{Geant4}}\xspace}

\def\photos     {\mbox{\textsc{Photos}}\xspace}

\def\tell1  {TELL1\xspace}
\def\ukl1   {UKL1\xspace}

\usepackage{mciteplus}

\newcommand{\ptrans}{\ensuremath{p_{\rm T}}}

\newcommand{\jpsimumu}{\ensuremath{\jpsi\to\mu^+ \mu^-}}

\newcommand{\bplus}{\ensuremath{B^{\rm +}}}
\newcommand{\bzero}{\ensuremath{B^{\rm 0}}}
\newcommand{\bszero}{\ensuremath{B^{\rm 0}_s}}

\newcommand{\bpjpsik}{\ensuremath{\bplus\to\jpsi K^+}}
\newcommand{\bzjpsikstar}{\ensuremath{\bzero\to\jpsi K^{*0}}}
\newcommand{\kstar}{\ensuremath{K^{*0}}}
\newcommand{\kstarkpi}{\ensuremath{\kstar\to K^+\pi^-}}

\newcommand{\bszjpsiphi}{\ensuremath{\bszero\to\jpsi \phi}}
\newcommand{\phikk}{\ensuremath{\phi\to K^+K^-}}
\newcommand{\cndof}{\ensuremath{\chi^2/{\rm ndf}}}
\newcommand{\bpyynobrresult}{\ensuremath{38.9\pm0.3\,({\rm stat.})\pm2.5\,({\rm syst.})\,\pm1.3\,({\rm norm.})\,{\rm \mub}}}
\newcommand{\bzresult}{\ensuremath{38.1\pm0.6\,({\rm stat.})\pm3.7\,({\rm syst.})\,\pm4.7\,({\rm norm.})\,{\rm \mub}}}
\newcommand{\bszresult}{\ensuremath{10.5\pm0.2\,({\rm stat.})\pm0.8\,({\rm syst.})\,\pm1.0\,({\rm norm.})\,{\rm \mub}}}

\newcommand{\bmeson}{$B$ meson}

\newcommand{\bmesons}{\bmeson s}

\newcommand{\lumivalue}{\ensuremath{362\pm13\invpb}}
\newcommand{\lumionly}{\ensuremath{0.36\invfb}}
\newcommand{\ppsymbol}{\ensuremath{pp}} 
\usepackage{longtable} 

\begin{document}

\renewcommand{\thefootnote}{\fnsymbol{footnote}}
\setcounter{footnote}{1}


\begin{titlepage}
\pagenumbering{roman}

\vspace*{-1.5cm}
\centerline{\large EUROPEAN ORGANIZATION FOR NUCLEAR RESEARCH (CERN)}
\vspace*{1.5cm}
\hspace*{-0.5cm}
\begin{tabular*}{\linewidth}{lc@{\extracolsep{\fill}}r}
\ifthenelse{\boolean{pdflatex}}
{\vspace*{-2.7cm}\mbox{\!\!\!\includegraphics[width=.14\textwidth]{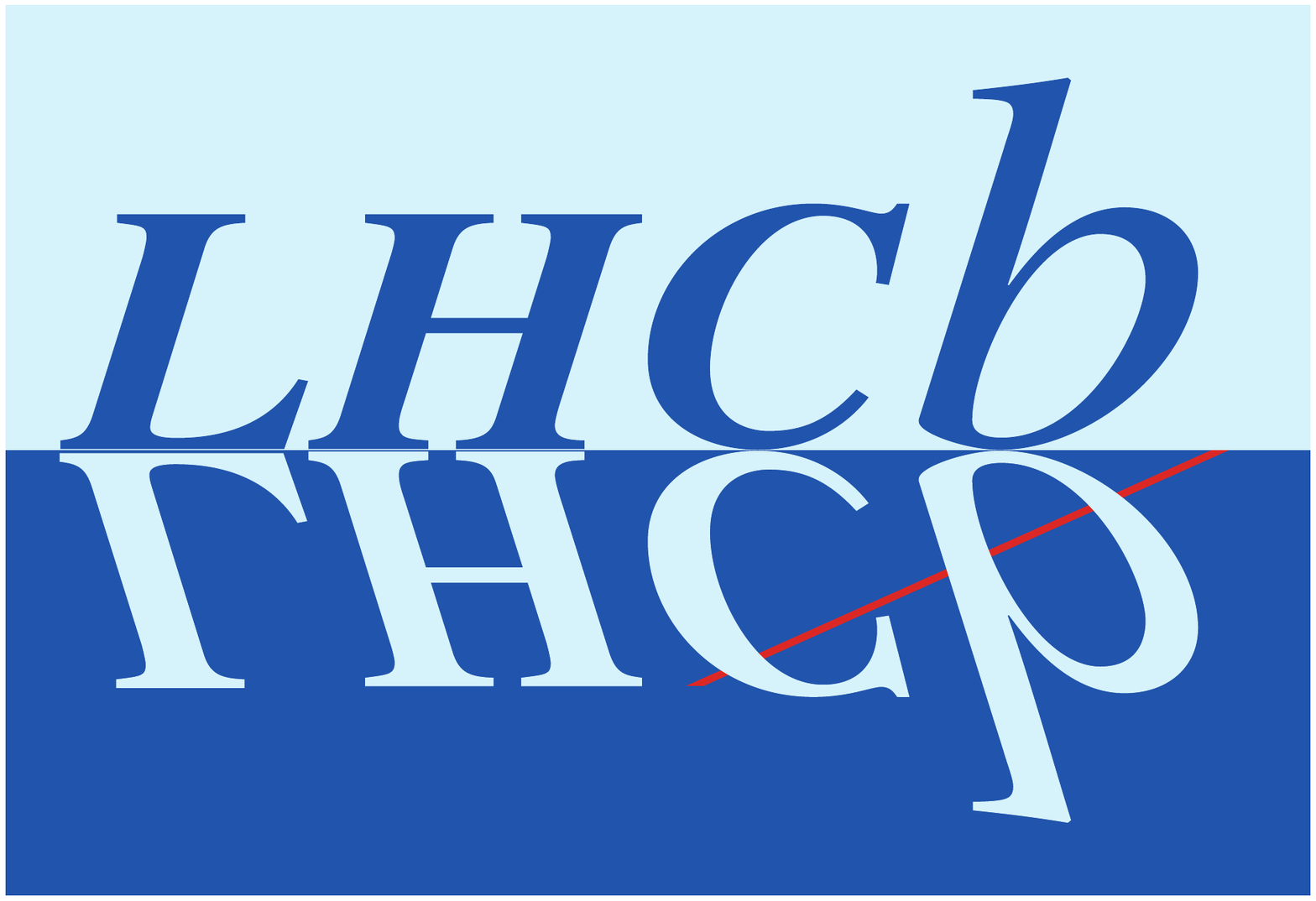}} & &}%
{\vspace*{-1.2cm}\mbox{\!\!\!\includegraphics[width=.12\textwidth]{figs/lhcb-logo.eps}} & &}%
\\
 & & CERN-PH-EP-2013-095 \\  
 & & LHCb-PAPER-2013-004 \\  
 & & Oct. 1, 2013 \\ 
 & & \\
\end{tabular*}

\vspace*{2.5cm}

{\bf\boldmath\huge
\begin{center}
Measurement of $B$ meson production cross-sections
in proton-proton collisions \mbox{at $\sqrt{s}=$ 7 TeV}
\end{center}
}

\vspace*{1.0cm}

\begin{center}
The LHCb collaboration\footnote{Authors are listed on the following pages.}
\end{center}

\vspace{\fill}

\begin{abstract}
  \noindent
The production cross-sections of $B$ mesons
are measured in $pp$ collisions at a centre-of-mass energy of $7\tev$,
using data collected with the LHCb detector corresponding to a
integrated luminosity of \lumionly.
The \bplus, \bzero\ and \bszero\ mesons
are reconstructed in the exclusive decays \bpjpsik,
\bzjpsikstar\ and \bszjpsiphi,
with \jpsimumu, $\kstarkpi$ and $\phi\to K^+K^-$.
The differential cross-sections are measured as
functions of $B$ meson transverse momentum \ptrans\ and rapidity $y$,
in the range $0<\ptrans<40\;\gevc$ and $2.0<y<4.5$.
The integrated cross-sections in the same \ptrans\ and $y$ ranges,
including charge-conjugate states,
are measured to be
\begin{equation}
\begin{array}{lcl}
\sigma(pp\to B^++X) &=& \bpyynobrresult,  \\
\sigma(pp\to B^0+X) &=& \bzresult,  \\
\sigma(pp\to B^0_s+X) &=& \bszresult,
\end{array}
\nonumber
\label{eq:result}
\end{equation}
where the third uncertainty arises from
the pre-existing branching fraction measurements.
\end{abstract}

\vspace*{1.0cm}
\begin{center}
  Submitted to JHEP
\end{center}

\vspace{\fill}

{\footnotesize
\centerline{\copyright~CERN on behalf of the \lhcb collaboration, license \href{http://creativecommons.org/licenses/by/3.0/}{CC-BY-3.0}.}}
\vspace*{2mm}

\end{titlepage}


\newpage
\setcounter{page}{2}
\mbox{~}
\newpage

\centerline{\large\bf LHCb collaboration}
\begin{flushleft}
\small
R.~Aaij$^{40}$, 
C.~Abellan~Beteta$^{35,n}$, 
B.~Adeva$^{36}$, 
M.~Adinolfi$^{45}$, 
C.~Adrover$^{6}$, 
A.~Affolder$^{51}$, 
Z.~Ajaltouni$^{5}$, 
J.~Albrecht$^{9}$, 
F.~Alessio$^{37}$, 
M.~Alexander$^{50}$
% ...

{\footnotesize \it
% ...
$ ^{5}$Clermont Universit\'{e}, Universit\'{e} Blaise Pascal, CNRS/IN2P3, LPC, Clermont-Ferrand, France\\
$ ^{6}$CPPM, Aix-Marseille Universit\'{e}, CNRS/IN2P3, Marseille, France\\
% ...
$ ^{9}$Fakult\"{a}t Physik, Technische Universit\"{a}t Dortmund, Dortmund, Germany\\
% ...
$ ^{35}$Universitat de Barcelona, Barcelona, Spain\\
$ ^{36}$Universidad de Santiago de Compostela, Santiago de Compostela, Spain\\
$ ^{37}$European Organization for Nuclear Research (CERN), Geneva, Switzerland\\
% ...
$ ^{40}$Nikhef National Institute for Subatomic Physics, Amsterdam, The Netherlands\\
% ...
$ ^{45}$H.H. Wills Physics Laboratory, University of Bristol, Bristol, United Kingdom\\
% ...
$ ^{50}$School of Physics and Astronomy, University of Glasgow, Glasgow, United Kingdom\\
$ ^{51}$Oliver Lodge Laboratory, University of Liverpool, Liverpool, United Kingdom\\
% ...
$ ^{n}$LIFAELS, La Salle, Universitat Ramon Llull, Barcelona, Spain\\
% ...
}
\end{flushleft}

\cleardoublepage


\renewcommand{\thefootnote}{\arabic{footnote}}
\setcounter{footnote}{0}



\pagestyle{plain} 
\setcounter{page}{1}
\pagenumbering{arabic}


%

\section{Introduction}
\label{sec:introduction}

Measurements of beauty production in multi-TeV proton-proton (\ppsymbol) collisions at the LHC
provide important tests of quantum chromodynamics.
State of the art theoretical predictions are given by the
fixed-order plus next-to-leading logarithm (FONLL)
approach~\cite{Cacciari:1998it,Cacciari:2001td,Cacciari:2012ny}.
In these calculations, the dominant uncertainties
arise from the choice of the renormalisation
and factorisation scales,
and the assumed \textit{b}-quark mass~\cite{Cacciari:2003uh}.
The primary products of \bbbar\ hadronisation are
\bplus, \bzero, \bszero\ and their charge-conjugate states
(throughout the paper referred to as \bmesons)
formed by one $\bar{b}$ quark bound to one of the three light quarks
($u$, $d$ and $s$).
Accurate measurements of the cross-sections
probe the validity of the production models.
At the LHC, \bbbar\ production has been studied
in inclusive
$b\to \jpsi X$ decays~\cite{Aaij:2011jh,Chatrchyan:2012hw}
and semileptonic~\cite{Aaij:2010gn,atlasbprod:2012fc} decays.
Other measurements, using fully reconstructed $B$ mesons,
have also been performed by the \lhcb
and CMS collaborations~\cite{Aaij:2012jd,Khachatryan:2011mk,Chatrchyan:2011pw,Chatrchyan:2011vh}.

In this paper, a measurement of
the production cross-sections of
\bmesons\ (including their charge-conjugate states)
is presented. This study is performed
in the transverse momentum range
$0<\ptrans<40\,\gevc$ and rapidity range $2.0<y<4.5$
using data, corresponding to a
integrated luminosity of \lumionly, collected
in \ppsymbol\ collisions at centre-of-mass energy of $7\tev$
by the LHCb experiment.
The $B$ mesons
are reconstructed in the exclusive decays \bpjpsik,
\bzjpsikstar\ and \bszjpsiphi,
with \jpsimumu, $\kstarkpi$ and $\phi\to K^+K^-$.

The \lhcb detector~\cite{Alves:2008td} is a single-arm forward
spectrometer covering the \mbox{pseudorapidity} range $2<\eta <5$,
designed for the study of particles containing \bquark or \cquark
quarks. The detector includes a high-precision tracking system
consisting of a silicon-strip vertex detector surrounding the \ppsymbol\
interaction region, a large-area silicon-strip detector located
upstream of a dipole magnet with a bending power of about
$4{\rm\,Tm}$, and three stations of silicon-strip detectors and straw
drift tubes placed downstream. The combined tracking system has
momentum resolution $\Delta p/p$ that varies from 0.4\% at 5\gevc to
0.6\% at 100\gevc, and impact parameter resolution 
better than 20\mum for transverse momentum higher than 3 \gevc.
Charged hadrons are identified
using two ring-imaging Cherenkov (RICH) detectors. Photon, electron and
hadron candidates are identified by a calorimeter system consisting of
scintillating-pad and preshower detectors, an electromagnetic
calorimeter and a hadronic calorimeter. Muons are identified by a
system composed of alternating layers of iron and multiwire
proportional chambers.

The events used in this analysis are selected
by a two-stage trigger system~\cite{Aaij:2012me}.
The first stage is hardware based whilst the second stage
is software based.
At the hardware stage events containing
either a single muon or a pair of muon candidates, with high transverse momentum, are selected.
In the subsequent software trigger
the decision of the single-muon or dimuon hardware trigger is confirmed and
a muon pair with an invariant mass consistent
with the known \jpsi\ mass~\cite{Nakamura:2012pd} is required.
To reject high-multiplicity events with a large number of \ppsymbol\ interactions,
global event cuts on the hit multiplicities of subdetectors are applied.

\section{Candidate selection}
\label{sec:eventselection}

The selection of \bmeson\ candidates starts by forming \jpsimumu\ decay candidates.
These are formed from pairs of
oppositely-charged particles
that are identified as muons and have $\ptrans>0.7\gevc$.
Good quality of the reconstructed tracks is ensured by requiring
the $\cndof$ of the track fit
to be less than 4, where ndf is the number of degrees
of freedom of the fit.
The muon candidates are required to originate
from a common vertex and the $\cndof$ of the vertex
fit is required to be less than 9.
The mass of the $\jpsi$ candidate is required to be
around the known \jpsi\ mass~\cite{Nakamura:2012pd},
between $3.04$ and $3.14\,\gevcc$.

Kaons used to form \bpjpsik\ candidates
are required to have \ptrans\ larger than 0.5 $\gevc$.
Information from the RICH detector system is not used in the selection
since the $\bplus\to\jpsi\pi^+$ decay is Cabibbo suppressed.
Candidates for \kstarkpi\ and \phikk\ decays
are formed from pairs of oppositely-charged hadron candidates.
Since the background levels of these two channels are higher
than for \bpjpsik decay, the hadron identification information
provided by the RICH detectors is used.
Kaons used to form \kstar\ candidates in the \bzjpsikstar\ channel
and $\phi$ candidates in the \bszjpsiphi\ channel
are selected by cutting on the difference between the log-likelihoods of the kaon and pion
hypotheses provided by the RICH detectors ($\dllkpi>0$).
The pions used to form \kstar\ candidates
are required to be inconsistent with the kaon hypothesis (${\rm DLL}_{\pi K}>-5$).
The same track quality cuts used for muons are applied to kaons and pions.
The \kstar\ and $\phi$ meson candidates are constructed
requiring a good vertex quality ($\cndof<16$) and $\ptrans>1.0\,\gevc$.
The masses of the \kstar\ and $\phi$ candidates
are required to be consistent with their known masses~\cite{Nakamura:2012pd},
in the intervals $0.826 - 0.966\,\gevcc$ and $1.008 - 1.032\,\gevcc$, respectively.

The $\jpsi$ candidate is combined with a \Kp, $K^{*0}$ or $\phi$ candidate
to form a \bplus, \bzero\ or \bszero\ meson, respectively.
A vertex fit~\cite{LHCb:dtf} is performed
that constrains the daughter particles
to originate from a common point and the mass of the muon pair to
match the known \jpsi\ mass~\cite{Nakamura:2012pd}.
The $\cndof$ returned by this fit is required to be less than 9.
To further reduce the combinatorial background due to particles produced
in the primary \ppsymbol\ interaction, only $B$ candidates
with a decay time larger than \mbox{0.3 ${\rm ps}$},
which corresponds to about 6 times the decay time resolution, are kept.
In the \bzjpsikstar\ samples, duplicate candidates are found
that share the same \jpsi\ particle but have pion tracks
that are reconstructed several times from one track.
In these cases only one of the candidates is randomly retained.
Duplicate candidates of other sources
in the other decay modes occur at a much lower rate and are retained.
Finally, the fiducial requirements \mbox{$0<\ptrans<40\;\gevc$}
and $2.0<y<4.5$ are applied to the $B$ meson candidates.

\section{Cross-section determination}
\label{sec:determination}

\noindent
The differential production cross-section for each $B$ meson species is calculated as
\begin{eqnarray}
\frac{{\rm d^2}\sigma(B)}{{\rm d}\ptrans\;{\rm d}y}&=&
\frac{N_{B}(\ptrans,y)}
{\epsilon_{\rm tot}(\ptrans,y)\;
{\cal L}_{\rm int}\;
{\cal B}(B\to\jpsi X)\;
\Delta\ptrans\;
\Delta y}, \nonumber
\label{eq:diffxsEq_11}
\end{eqnarray}
where $N_{B}(\ptrans,y)$ is the number of reconstructed signal candidates
in a given $(\ptrans,y)$ bin,
$\epsilon_{\rm tot}(\ptrans,y)$ is the total efficiency
in a given $(\ptrans,y)$ bin,
${\cal L}_{\rm int}$ is the integrated luminosity,
${\cal B}(B\to\jpsi X)$ is the product of
the branching fractions of the decays in the complete decay chain,
and $\Delta\ptrans$ and $\Delta y$ are the widths of the bin.
The width of each $y$ bin is fixed to 0.5
while the widths of the \ptrans\ bins vary to allow
for sufficient number of candidates in each bin.

\begin{figure}[t]
  \centering
  \includegraphics[width=7.5cm]{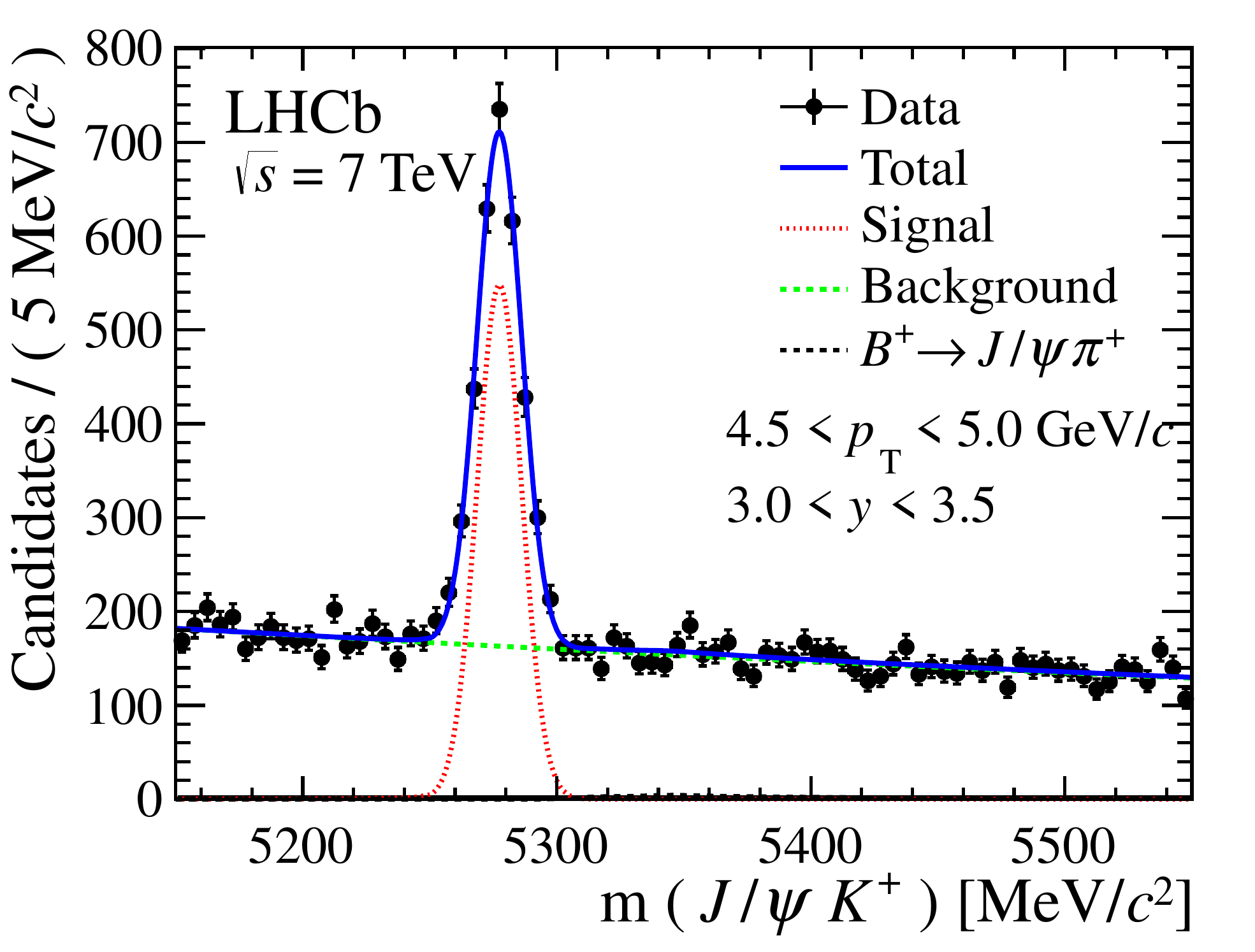}
    \includegraphics[width=7.5cm]{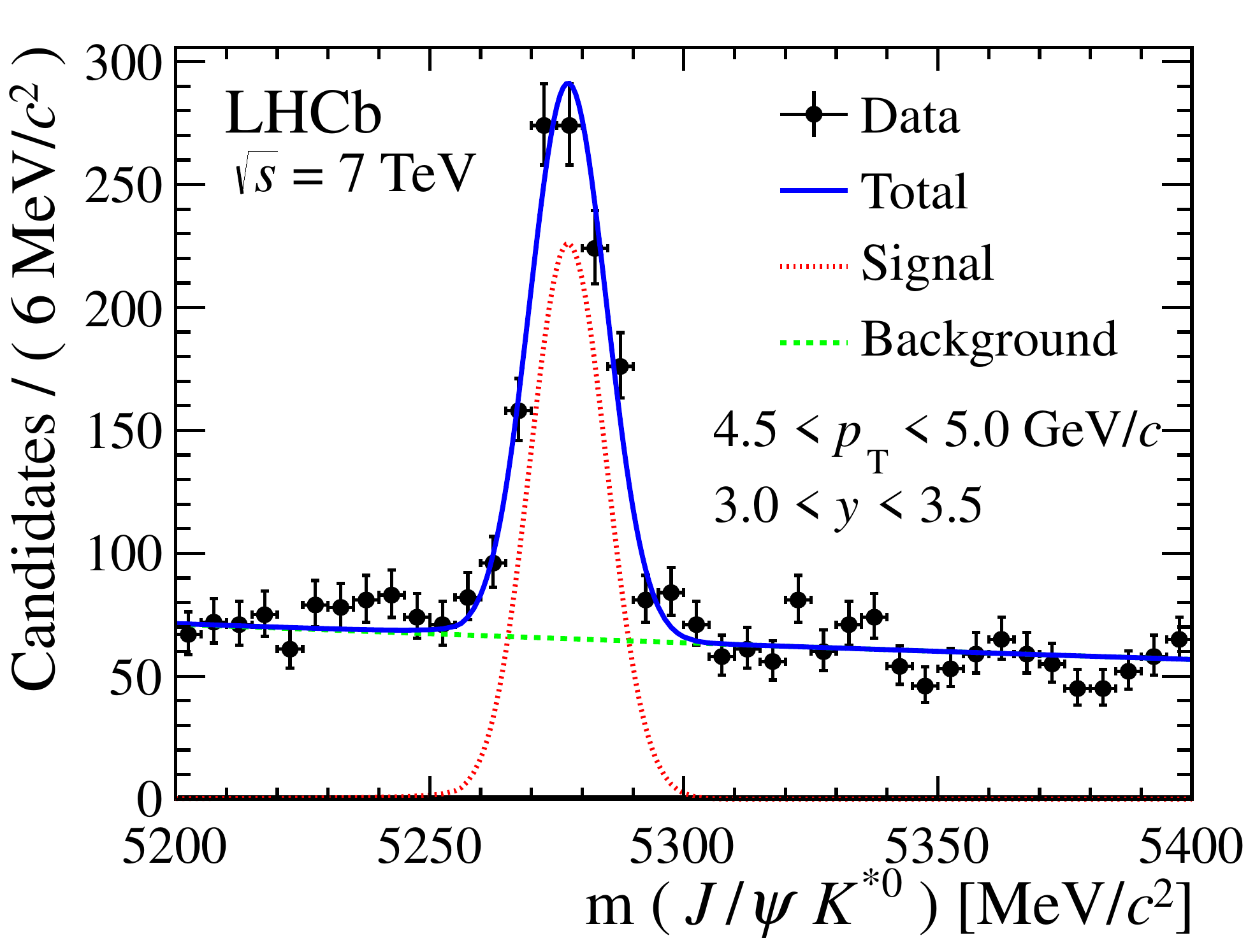}
    \includegraphics[width=7.5cm]{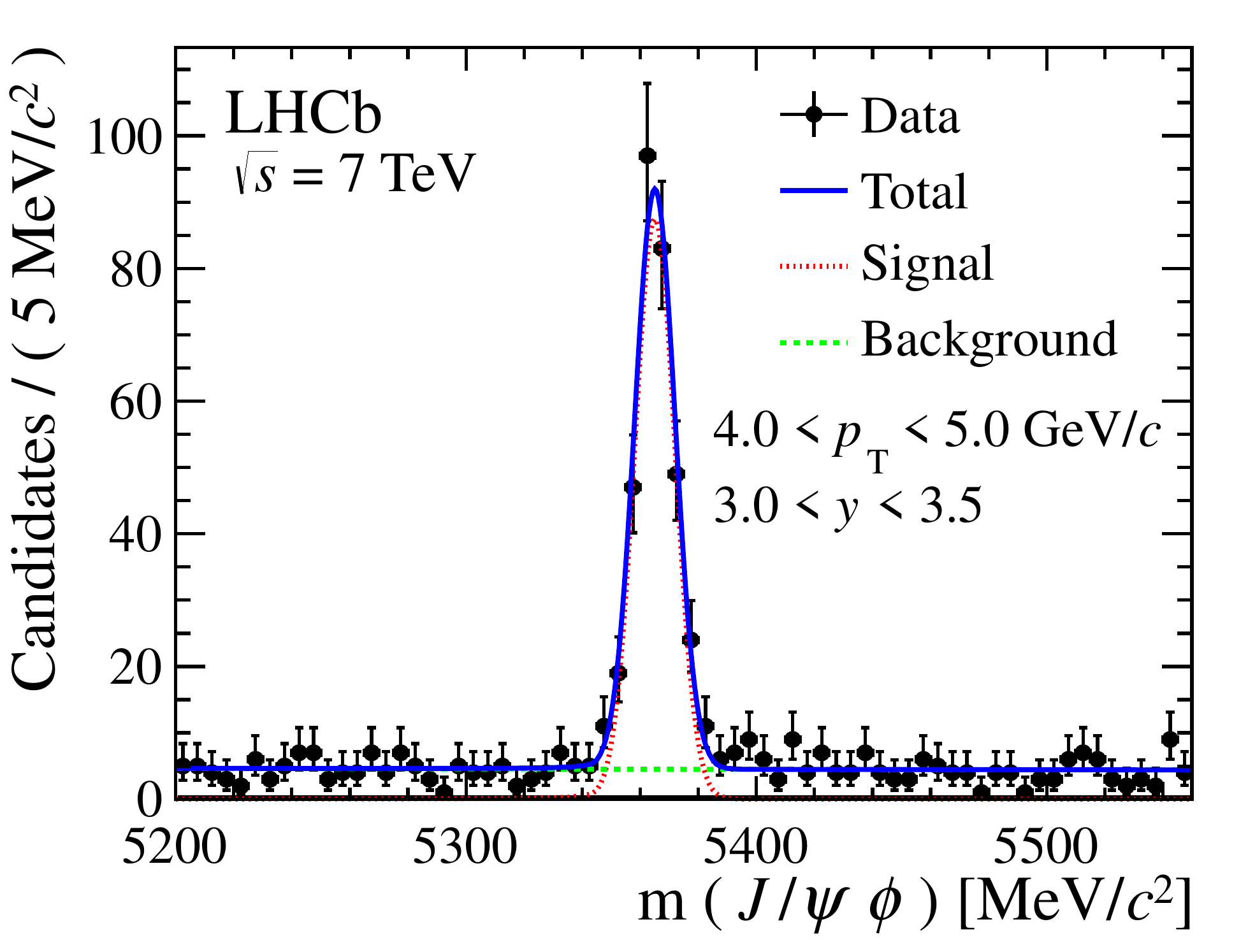}
  \caption{\small Invariant mass distributions of the selected
    candidates for ({\it top left}) \bplus\ and
    ({\it top right}) \bzero\ decays, both in the range
    $4.5<\ptrans<5.0\;\gevc$, $3.0<y<3.5$,
    and ({\it bottom}) \bszero\ decay in the range
    $4.0<\ptrans<5.0\;\gevc$, $3.0<y<3.5$.
    The results of the fit to the model described in the text
    are superimposed.
    The Cabibbo-suppressed background is barely visible
    in the top left plot.
  }
  \label{fig:BuMassPlot}
\end{figure}

The signal yield in each bin of \ptrans\ and $y$ is determined
using an extended unbinned maximum likelihood fit to the invariant mass
distribution of the reconstructed $B$ candidates.
The fit model includes two components: a double-sided Crystal Ball function to
model the signal and an exponential function to model the combinatorial background. The former
is an extension of the Crystal Ball function~\cite{Skwarnicki:1986ps}
that has tails on both the low- and the high-mass side
of the peak described by separate parameters,
which are determined from simulation.
For the $\bplus$ channel, the $K$-$\pi$ misidentified $\bplus\to\jpsi\pi^+$ decay
is modelled by a shape that is found to
fit the distribution of simulated events.
The invariant mass distributions of the selected $B$ candidates
and the fit results in one \ptrans\ and $y$ bin are shown
in Fig.~\ref{fig:BuMassPlot}.

For the $\bzjpsikstar$ and $\bszjpsiphi$
decay channels, an additional non-resonant S-wave component
(where the $K^+\pi^-$ and the $K^+K^-$ originate directly
from \bzero\ and \bszero\ decays, and not via $\kstar$ or
$\phi$ resonances) is also present.
The amount of this component present in each case is determined from
an independent fit to the $K^+\pi^-$ or $K^+K^-$ mass distribution, respectively,
integrating the $\ptrans$ and $y$ range.
The signal component is described by a relativistic Breit-Wigner function,
and the S-wave background by a phase space function.
From the fit results, the S-wave fractions 
are determined to be $\sim6\%$ for $\bzero\to\jpsi K^+\pi^-$ and $\sim3\%$ for
$\bszero\to\jpsi K^+K^-$ decays.
The yields of $B$ mesons
are then corrected according to the S-wave fractions.

The geometrical acceptance as well as the reconstruction and selection
efficiencies, except for the hadron identification efficiencies,
are determined using simulated signal events.
The \ppsymbol\ collisions are generated using
\pythia~6.4~\cite{Sjostrand:2006za} with a specific \lhcb\
configuration~\cite{LHCb-PROC-2010-056}.  Decays of hadronic particles
are described by \evtgen~\cite{Lange:2001uf}, in which final state
radiation is generated using \photos~\cite{Golonka:2005pn}. The
interaction of the generated particles with the detector and its
response are implemented using the \geant
toolkit~\cite{Allison:2006ve, *Agostinelli:2002hh} as described in
Ref.~\cite{LHCb-PROC-2011-006}.
The hadron identification efficiencies
are measured using tracks
from the decay $D^{*+}\to D^0\pi^+$ with $D^0 \to K^-\pi^+$,
selected without using
information from the RICH detectors~\cite{LHCb-DP-2012-003}.

Only candidates where the \jpsi\ is responsible
for the trigger decision are used.
The trigger efficiency
is measured in data using the tag and probe method described in Ref.~\cite{Aaij:2012me}.

The luminosity is measured using Van der Meer scans
and a beam-gas imaging method~\cite{Aaij:2011er}.
The integrated luminosity of the data sample used in this analysis
is determined to be \lumivalue.

The branching fraction ${\cal B}(\bzjpsikstar)$ measured by the \belle\ collaboration~\cite{Abe:2002haa}
is used in the determination of the \bzero\ cross-section,
since it includes the effect of the S-wave interference, while other measurements do not.
The measurement of ${\cal B}(\bszjpsiphi)$
given in Ref.~\cite{Aaij:2013orb}
is used in the determination of the \bszero\ cross-section.
Since this branching fraction measurement used the average
ratio of fragmentation fractions
$f_s/f_d$ from Ref.~\cite{Aaij:2011jp,Aaij:2013qqa},
the result in this paper cannot be taken as an independent measurement of $f_s/f_d$.
The other branching fractions are obtained from Ref.~\cite{Nakamura:2012pd}.

\section{Systematic uncertainties}
\label{sec:determination}
The measurements are affected by systematic uncertainties in the
determination of the signal yields, efficiencies, branching fractions
and luminosity, as summarised in \mbox{Table~\ref{tab:syst11}}.
The total systematic uncertainty is obtained from the sum in quadrature
of all contributions.

\renewcommand{\arraystretch}{1.2}
\begin{table}[t]
\tabcolsep 4mm
\begin{center}
\caption{\small \label{tab:syst11}Relative systematic uncertainties (in \%),
given as single values or as ranges, when they depend
on the (\ptrans, y) bin.}
  \scalebox{1}{%
\begin{tabular}{llll}
\toprule
Source  & \bplus  & \bzero\  & \bszero\  \\
\midrule
Signal fit model  &  2.5 & 1.3 & 1.2  \\
Fit range & 0.1 & 1.0 &  0.4\\
Non-resonant background & - & 2.2 & 1.9 \\
Combinatorial background & 0.6 & 0.8 & 0.3 \\
Bin size    &  $0.1-10.9$ & $0.1-19.3$ &  $0.1-13.2$ \\
Duplicate candidates & - & 3.1 & - \\
\midrule
Trigger efficiency  &   $2.4-7.9$ &   $2.6-7.9$&   $2.6-6.5$ \\
Tracking efficiency &  $2.4-7.4$ &  $4.4-8.3$ & $4.4-8.5$ \\
Vertex quality cut & 1.0& 0.9  & 0.2 \\
Muon identification & $0.7-4.9$ & $0.8-5.0$ & $0.8-5.8$ \\
Hadron identification & - & 1.0 &  0.8\\
Angular distribution & - & $0.1-0.3$ & $0.1-4.7$\\
\ptrans\ distribution & - & $0-24.4$ & - \\
\midrule
Branching fractions  & 3.3 & 12.3  & 10.0 \\
Luminosity  &  3.5& 3.5 &3.5 \\
\bottomrule
\end{tabular}}
\end{center}
\end{table}
\renewcommand{\arraystretch}{1.}

Uncertainties on the signal yields
arise from imperfect knowledge of the signal shape, non-resonant background and
finite size of the bins.
The uncertainty from the signal shape
is estimated by comparing the fitted and generated signal yields in simulation.
The non-resonant background ratios determined in this analysis are compared with
those from measurements with angular fits~\cite{Aaij:2013orb}
and the differences are assigned as systematic uncertainties.
By varying the \ptrans\ or $y$ binning,
the uncertainty for changing the binning in \ptrans\ is found to be small
while that for $y$ is non-negligible in the low $y$ bin.
An uncertainty is
assigned due to the procedure of removal of duplicate candidates
in \bzjpsikstar\ events. For the other modes this effect is found to be negligible.
The uncertainties from the background shape,
misidentified $\bplus\to\jpsi\pi^+$ background
and mass fit range are small.

Uncertainties on the efficiencies
arise from the trigger, tracking,
particle identification, angular distribution, \ptrans\ spectrum
and vertex fit quality cut.
The systematic uncertainty from the trigger efficiency
is evaluated by comparing the efficiency measured
using a trigger-unbiased sample of simulated \jpsi\ events
with that determined from the simulation.
The effect of the global event cuts in the trigger is found to be negligible.
The tracking efficiencies are
estimated with a tag and probe method~\cite{Jaeger:1402577}
using $\jpsi\rightarrow\mu^+\mu^-$ events in both data and simulation.
The simulated efficiencies, used to determine the cross-section,
are corrected according to the differences between data and simulation.
The tracking uncertainty includes two components: the first
is from the data-simulation difference correction;
the second is due to the uncertainty
on the hadronic interaction length
of the detector used in the simulation.
Possible systematic biases in the determination of the hadron identification
efficiency are estimated using simulated events and comparing the
true efficiency with that obtained by applying the same procedure
as for the data.
The muon identification uncertainty is estimated by
comparing the efficiency in simulation with
that measured, on data, using a tag and probe method.
The systematic uncertainties due to the uncertainties
on the angular distributions of \bzero\
and $\bszero$ decays~\cite{Nakamura:2012pd,LHCb:2011aa} are taken into
account by simulating the effect of varying the central values
of the polarization amplitudes by $\pm1$ sigma.
In the first \ptrans\ bin of the \bzero\ sample,
the agreement of the \ptrans\ distributions between data and simulation
is not as good as in the other bins.
The discrepancy is assigned as an additional uncertainty for that bin.
The vertex fit quality cut uncertainty is estimated from the
data to simulation comparison.
By calculating the signal yields and efficiencies separately
for data taken with two magnet polarities,
the results are found to be stable.

The systematic uncertainties from the branching fractions are calculated
with their correlations taken into account.
Since the ${\cal B}(\bzjpsikstar)$ and ${\cal B}(\bszjpsiphi)$
have been measured with sizeable uncertainty,
the corresponding uncertainties are listed separately in the cross-section results.
The absolute luminosity scale is measured
with $3.5\%$ uncertainty,
which is dominated by the beam current uncertainty~\cite{Aaij:2011er}.

\section{Results and conclusion}
\label{sec:results}
\begin{figure}
\centering
\includegraphics[width=8.5cm]{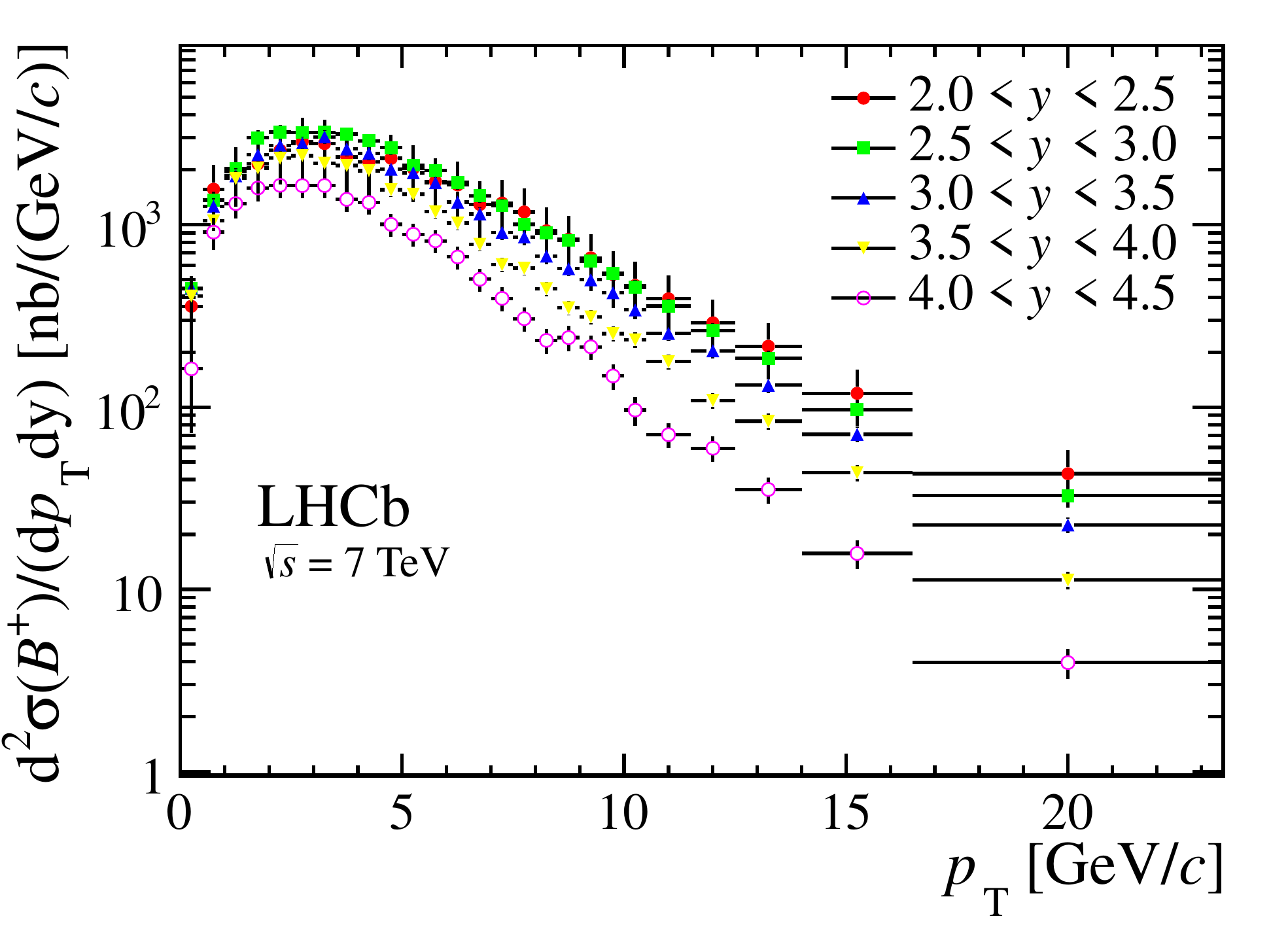}
\includegraphics[width=8.5cm]{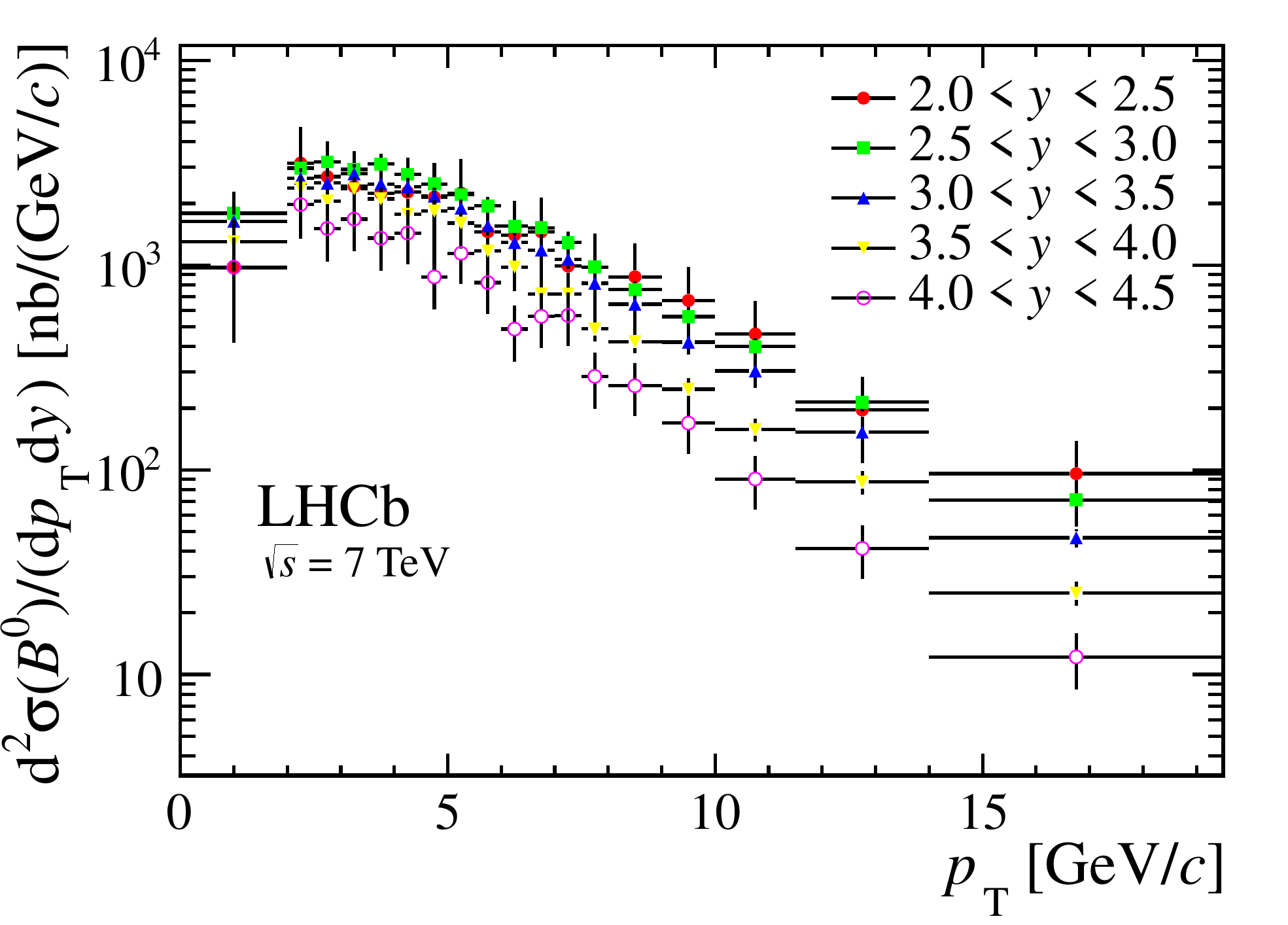}
\includegraphics[width=8.5cm]{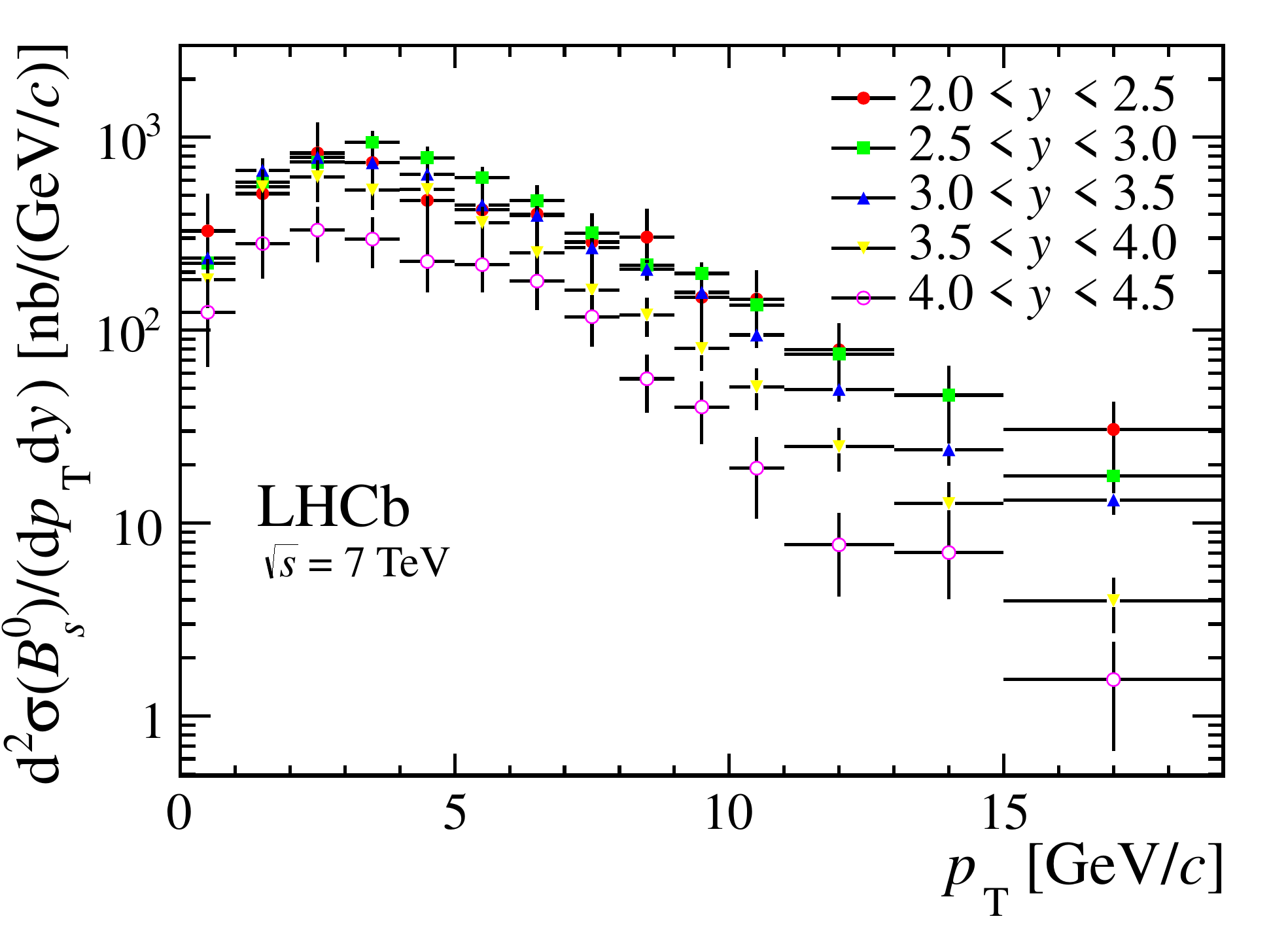}
\caption{\small Differential production cross-sections for
({\it top}) \bplus, ({\it middle}) \bzero\ and ({\it bottom}) \bszero\
mesons, as functions of \ptrans\
for each $y$ interval.}
\label{fig:bu_2Dresults}
\end{figure}

\begin{figure}
\centering
\includegraphics[width=9cm]{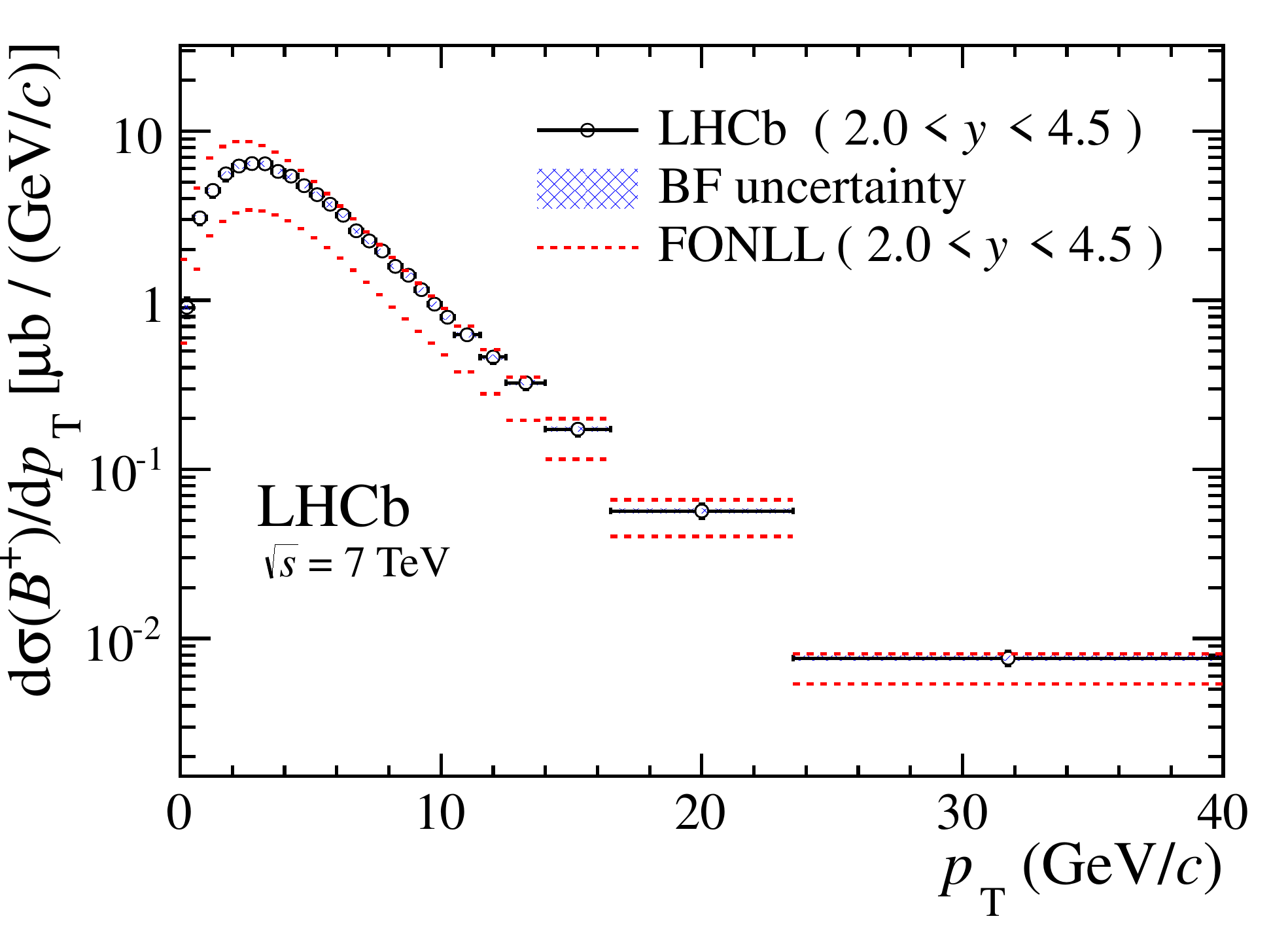}
\includegraphics[width=9cm]{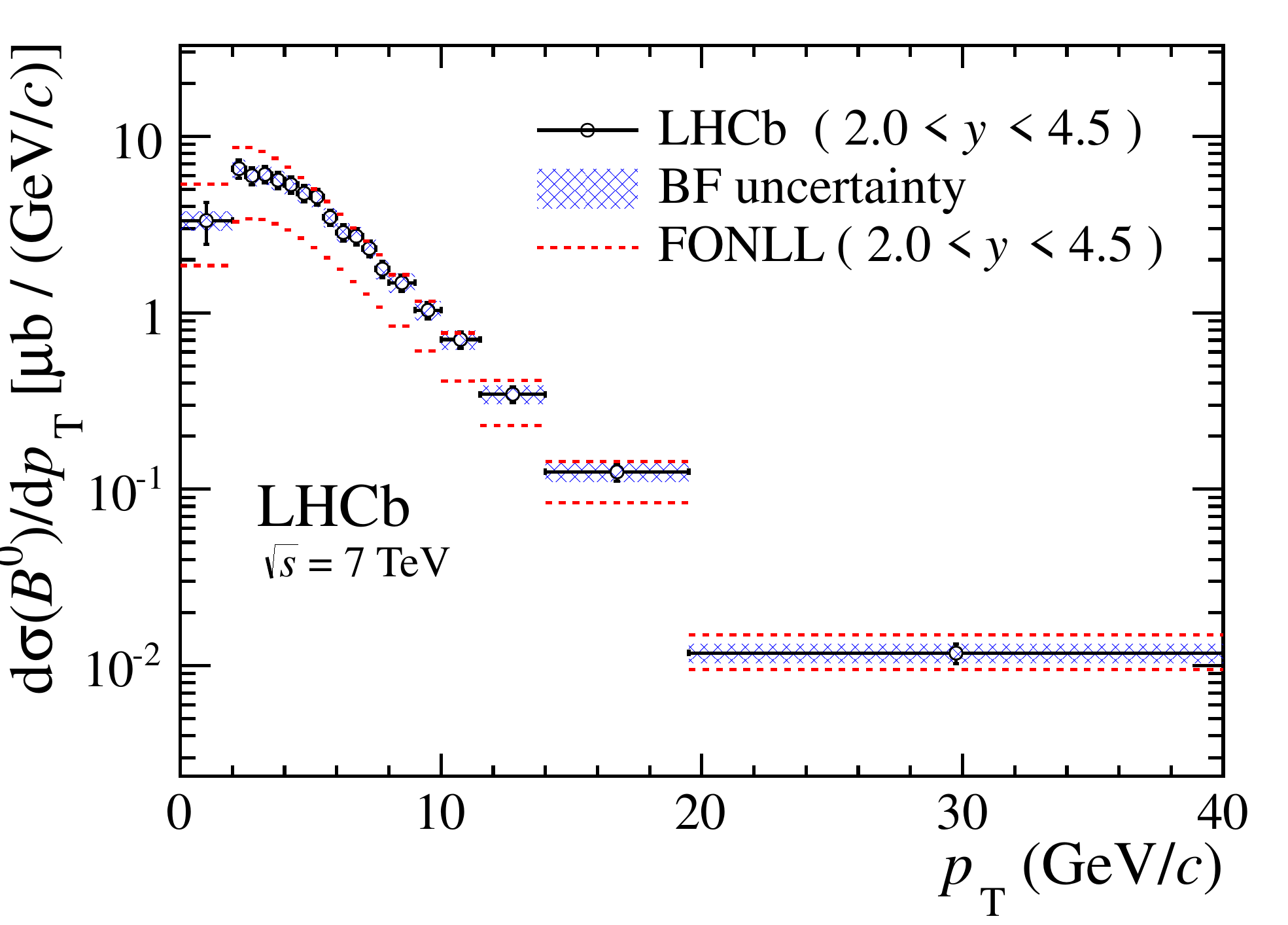}
\includegraphics[width=9cm]{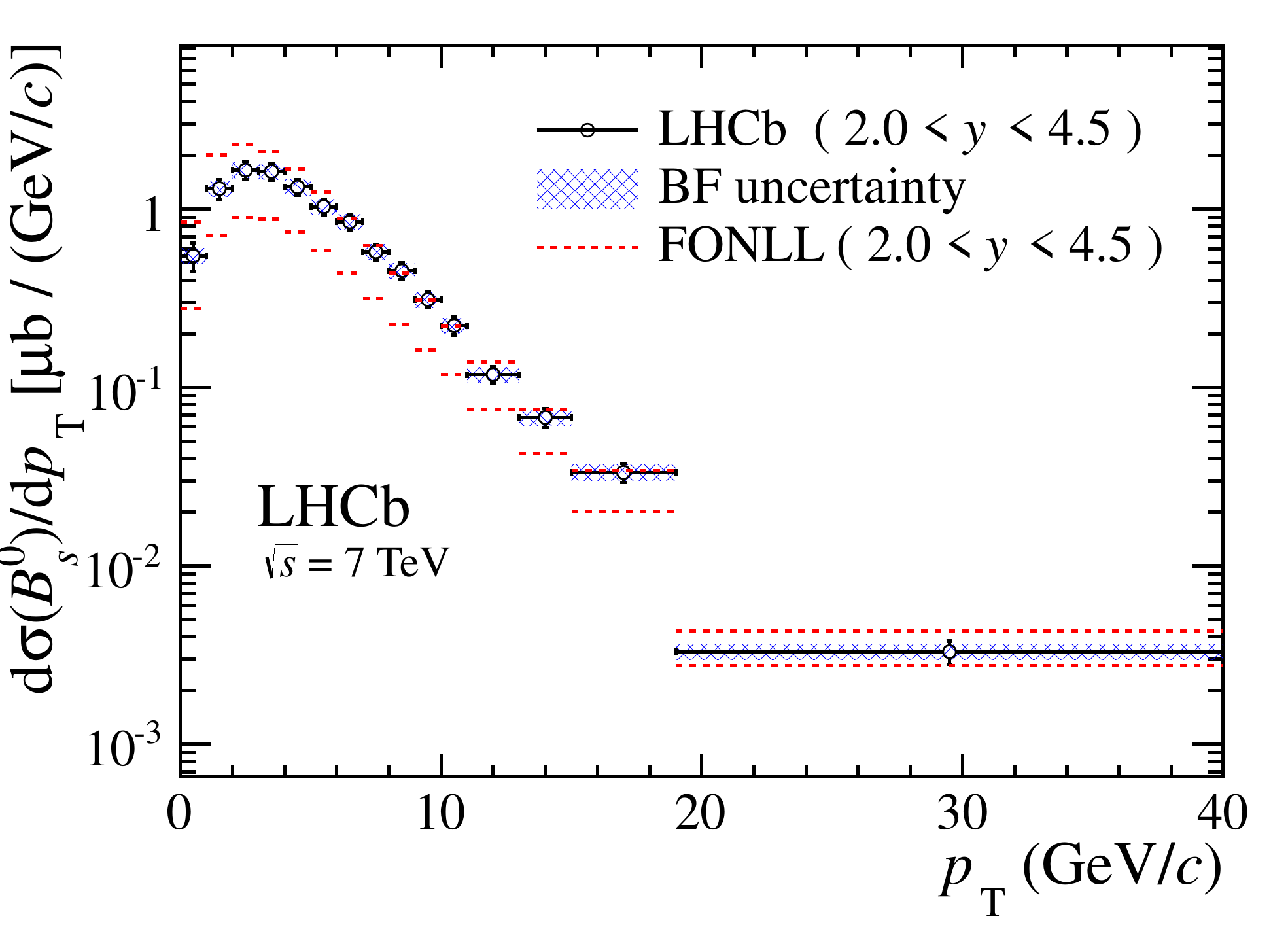}
\caption{\small Differential production cross-sections for
({\it top}) \bplus, ({\it middle}) \bzero\ and ({\it bottom}) \bszero\
mesons, as functions of \ptrans\
integrated over the whole $y$ range.
The open circles with error bars are
the measurements (not including
uncertainties from normalisation channel branching fractions and luminosity)
and the blue shaded areas are
 the uncertainties from the branching fractions.
 The red dashed lines
  are the upper and lower uncertainty limits of the \textsc{FONLL} computation~\cite{Cacciari:2012ny}.}
\label{fig:bu_results}
\end{figure}

\begin{figure}[t]
\centering
\includegraphics[width=9cm]{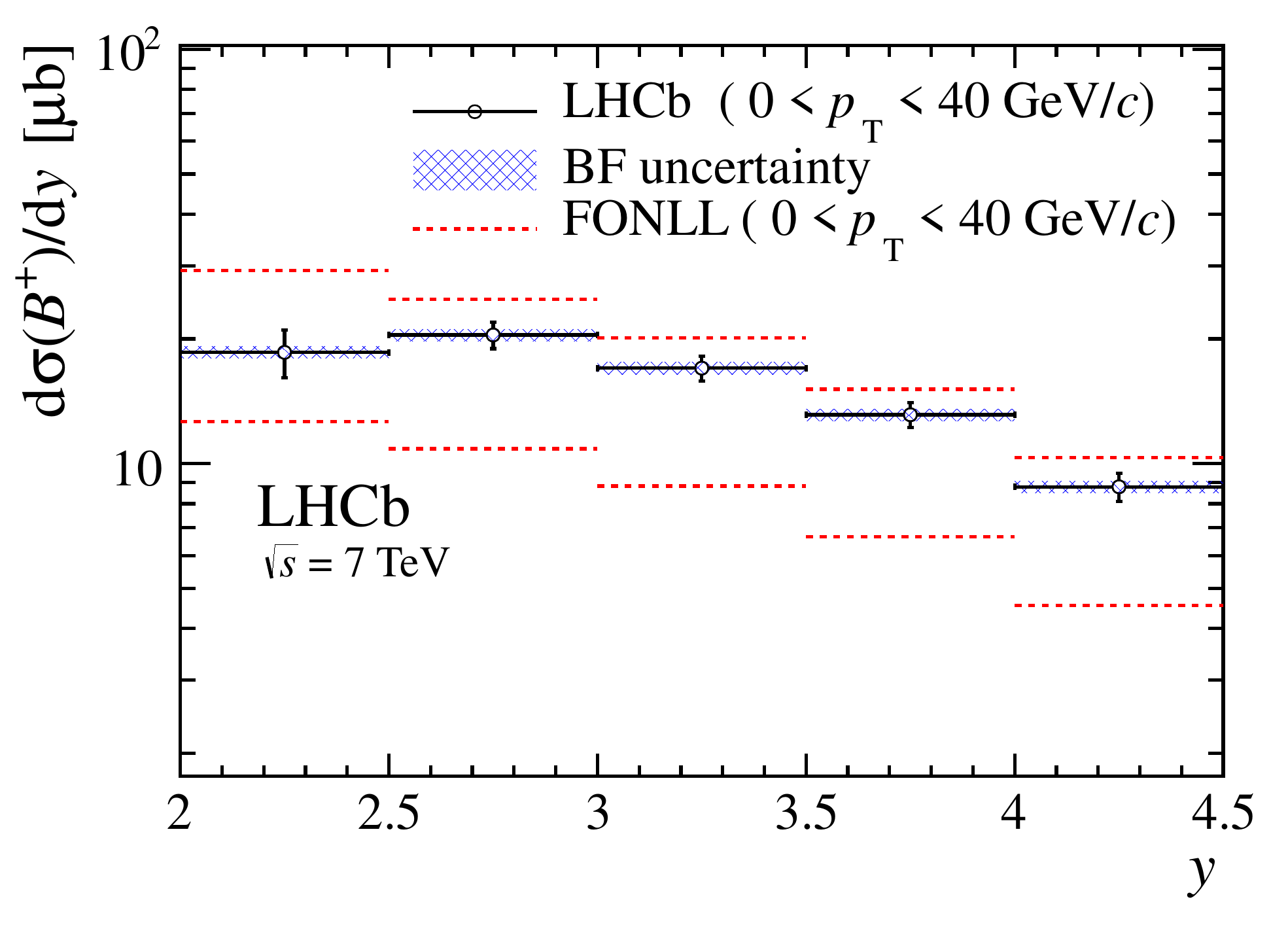}
\includegraphics[width=9cm]{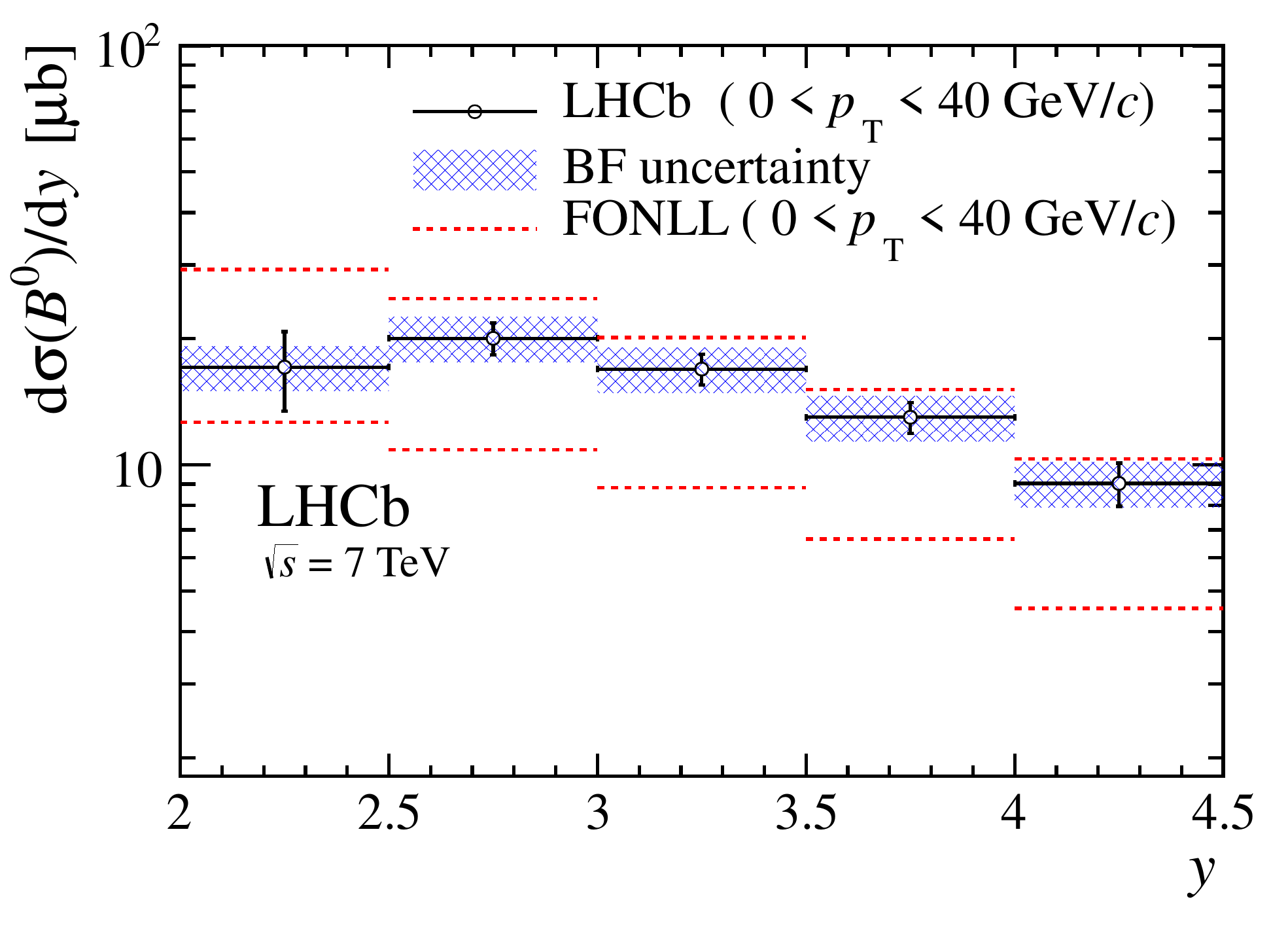}
\includegraphics[width=9cm]{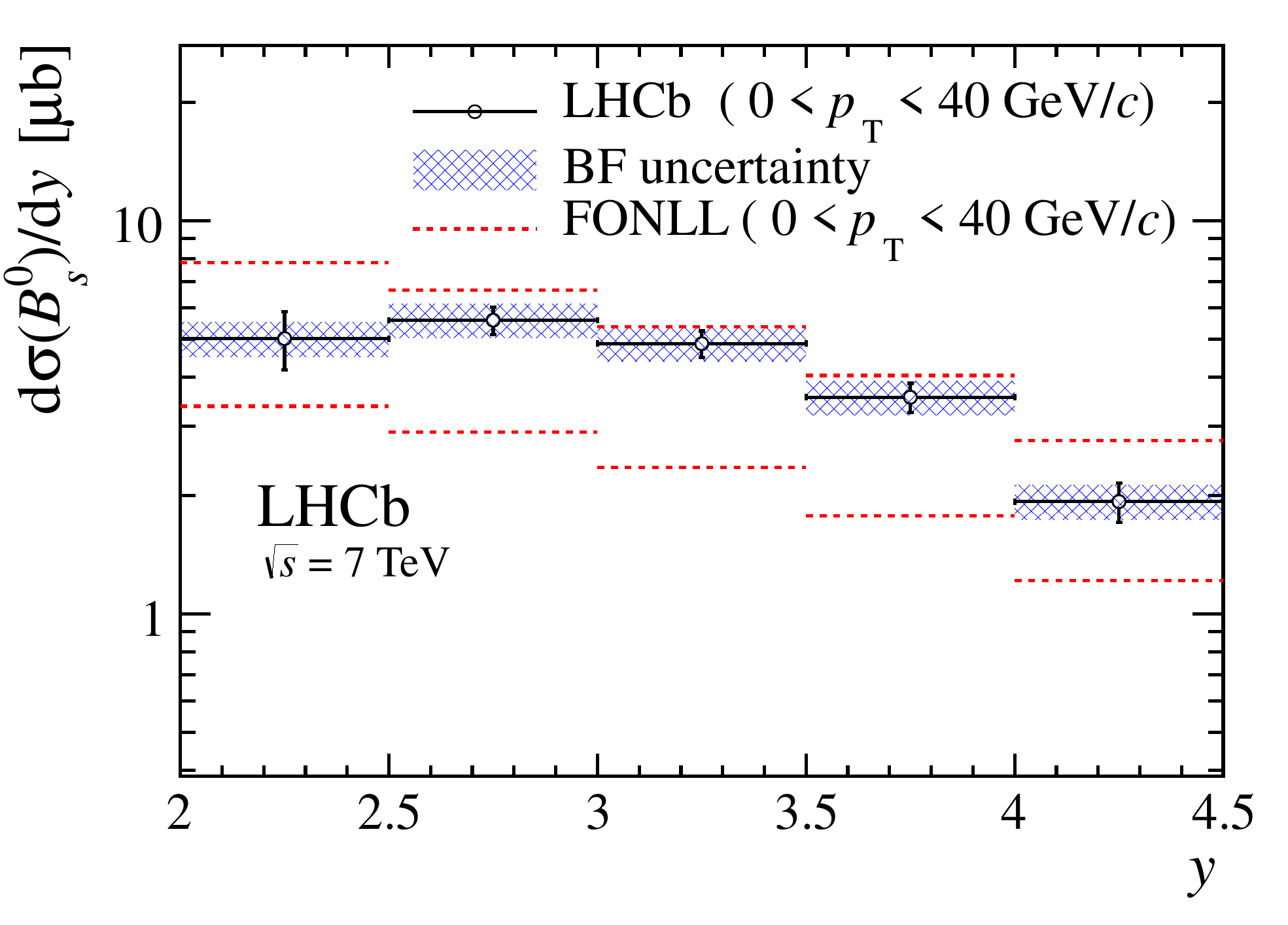}
\caption{\small Differential production cross-sections for
({\it top}) \bplus, ({\it middle}) \bzero\ and ({\it bottom}) \bszero\
mesons, as functions of $y$
integrated over the whole \ptrans\ range.
The black open circles with error bars are
the measurements (not including
uncertainties from normalisation channel branching fractions and luminosity)
and the blue shaded areas are
the uncertainties from the branching fractions.
The red dashed lines
are the upper and lower uncertainty limits of the \textsc{FONLL}
computation~\cite{Cacciari:2012ny}.}
\label{fig:yfonll}
\end{figure}

The measured differential production cross-sections of \bmesons\
in bins of \ptrans\ and $y$ are shown
in Fig.~\ref{fig:bu_2Dresults}.
These results are integrated separately over $y$ and \ptrans, and compared with
the \textsc{FONLL} predictions~\cite{Cacciari:2012ny},
as shown in Fig.~\ref{fig:bu_results}
and Fig.~\ref{fig:yfonll}, respectively.
The hadronisation fractions $f_u=f_d=(33.7\pm2.2)$\%
and $f_s=(9.0\pm0.9)$\% from Ref.~\cite{Aaij:2011jp}
are used to fix the overall scale of \textsc{FONLL}.
The uncertainty of the \textsc{FONLL} computation includes the uncertainties on the $b$-quark mass,
renormalisation and factorisation scales, and CTEQ~6.6~\cite{Nadolsky:2008zw}
parton distribution functions.
Good agreement is seen between the \textsc{FONLL} predictions
and measured data.

The integrated cross-sections of the
$B$ mesons with $0<\ptrans<40\;\gevc$ and $2.0<y<4.5$ are
\begin{equation}
\begin{array}{rcl}
\sigma(pp\to B^+\,X) &=& \bpyynobrresult,  \\
\sigma(pp\to B^0\,X) &=& \bzresult,  \\
\sigma(pp\to B^0_s\,X) &=& \bszresult,
\end{array}
\nonumber
\label{eq:result}
\end{equation}
where the third uncertainties arise from
the uncertainties of the branching fractions used for normalisation.
The \bplus\ result is in good agreement with a previous measurement
by \lhcb~\cite{Aaij:2012jd}.
These represent the first measurements of \bzero\
and \bszero\ meson production cross-sections in $pp$ collisions in the forward region at
centre-of-mass energy of $7\tev$.

\section*{Acknowledgements}

\noindent
We thank M. Cacciari for providing the FONLL predictions 
for the \bmeson\ production cross-sections.
We express our gratitude to our colleagues in the CERN
accelerator departments for the excellent performance of the LHC. We
thank the technical and administrative staff at the LHCb
institutes. We acknowledge support from CERN and from the national
agencies: CAPES, CNPq, FAPERJ and FINEP (Brazil); NSFC (China);
CNRS/IN2P3 and Region Auvergne (France); BMBF, DFG, HGF and MPG
(Germany); SFI (Ireland); INFN (Italy); FOM and NWO (The Netherlands);
SCSR (Poland); ANCS/IFA (Romania); MinES, Rosatom, RFBR and NRC
``Kurchatov Institute'' (Russia); MinECo, XuntaGal and GENCAT (Spain);
SNSF and SER (Switzerland); NAS Ukraine (Ukraine); STFC (United
Kingdom); NSF (USA). We also acknowledge the support received from the
ERC under FP7. The Tier1 computing centres are supported by IN2P3
(France), KIT and BMBF (Germany), INFN (Italy), NWO and SURF (The
Netherlands), PIC (Spain), GridPP (United Kingdom). We are thankful
for the computing resources put at our disposal by Yandex LLC
(Russia), as well as to the communities behind the multiple open
source software packages that we depend on. 
\clearpage

\addcontentsline{toc}{section}{References}
\setboolean{inbibliography}{true}
\bibliographystyle{LHCb}
\bibliography{main,LHCb-PAPER,LHCb-CONF,LHCb-DP}

\end{document}